\documentclass[sigconf,nonacm]{acmart}

\usepackage{booktabs}
\usepackage{graphicx}
\usepackage{enumitem}
\usepackage{tabularx}
\usepackage{afterpage}
\usepackage{float}
\newcommand{\ident}[1]{\texttt{\footnotesize #1}}
\setcopyright{none}
\acmConference[]{}{}{}
\renewcommand\footnotetextcopyrightpermission[1]{}

\AtBeginDocument{%
  \setlength{\textfloatsep}{8pt plus 0pt minus 2pt}%
  \setlength{\floatsep}{8pt plus 0pt minus 2pt}%
  \setlength{\intextsep}{8pt plus 0pt minus 2pt}%
}

\title{Do Vision-Language Models Agree on the Affective Qualities of Shape? A Cross-Model Audit for Generative Design Interfaces}

\author{Luca Bux}
\orcid{0000-0000-0000-0000}
\affiliation{
  \institution{Honda Research Institute Europe GmbH}
  \city{Offenbach am Main}
  \country{Germany}}
\email{luca.bux@honda-ri.de}

\author{Thiago Rios}
\orcid{0000-0000-0000-0000}
\affiliation{
  \institution{Honda Research Institute Europe GmbH}
  \city{Offenbach am Main}
  \country{Germany}}
\email{thiago.rios@honda-ri.de}

\author{Ingo Scholtes}
\affiliation{
  \institution{Julius-Maximilians-Universität Würzburg}
  \city{Würzburg}
  \country{Germany}}
\email{ingo.scholtes@uni-wuerzburg.de}

\author{Stefan Menzel}
\affiliation{
  \institution{Honda Research Institute Europe GmbH}
  \city{Offenbach am Main}
  \country{Germany}}
\email{stefan.menzel@honda-ri.de}

\author{Bernhard Sendhoff}
\affiliation{
  \institution{Honda Research Institute Europe GmbH}
  \city{Offenbach am Main}
  \country{Germany}}
\email{bernhard.sendhoff@honda-ri.de}

\begin{document}
\begin{abstract}
Generative design interfaces increasingly expose semantic controls that let users steer output with concepts such as ``more elegant'' or ``more minimalist,'' typically encoded by a vision-language model (VLM). A practical question is whether state-of-the-art VLMs represent objects consistently in terms of the same concept. We audit 6 VLMs by ranking untextured 3D objects along Kansei adjective pairs, where Kansei describes affective impressions of product form, with each axis defined as the difference between the text representations of its two poles. Geometric pairs serve as positive controls, and pairs of unrelated adjectives establish an empirical null. Across 10 categories of ShapeNet database, affective axes converge above the null (mean pairwise rank correlation $0.36$ vs.\ $0.14$) but below the geometric ceiling ($0.44$). The agreement between models is partial and highly uneven: on the three axes shared by all categories, mean convergence ranges from $0.21$ for bookshelves to $0.51$ for jars. Convergence depends primarily on whether a category's representational variation aligns with the semantic direction being evaluated, rather than simply on how much the objects vary in shape overall. Cross-model convergence does not imply agreement with human judgments. Based on our findings, we implement a UI prototype that shows how the audit can inform which Kansei descriptors to expose as controls for a given object class and which to withhold.
\end{abstract}

\maketitle
\section{Introduction}
Deep generative models have made it easier to explore many design alternatives during early-stage design~\cite{regenwetter2022deep}. Recent multimodal systems extend this process by transforming sketches or text prompts into visual concepts~\cite{song2024multimodal}, while text-to-3D methods generate three-dimensional objects directly from natural language~\cite{poole2023dreamfusion,lin2023magic3d}. Designers increasingly interact with these systems through semantic requests such as ``a sleeker handle,'' ``a more elegant chair,'' or ``a minimalist silhouette''~\cite{parikh2011relative,yumer2015semantic}. Such descriptions function as interface controls that guide semantic editing, optimization, or iterative refinement of designs. Before exposing these controls to users, it is important to characterize whether they correspond to stable semantic signals or reflect the behavior of individual models.

Vision-language models (VLMs) are increasingly used in AI-assisted design workflows to evaluate generated objects, guide optimization, and provide feedback during iterative design~\cite{wong2024promptevolution,radford2021clip}. Most of the existing work focuses on object recognition, \textit{e.g.}, verifying if a generated object is recognized as a chair, bottle, or car~\cite{radford2021clip}. Much less is known about whether VLMs consistently represent \textit{affective} qualities expressed through shape. Following Kansei engineering, which studies how properties of products and their forms relate to people's affective impressions~\cite{nagamachi1995kansei}, we use \textit{affective} to describe qualities of form that influence how people interpret products. This raises a practical question for generative design interfaces: \textit{Can we identify affective semantic controls that are sufficiently consistent across VLMs before exposing them in generative design systems?}

Evaluating this question directly against human perception is difficult at scale. Large annotated datasets of affective 3D shapes are limited, and collecting reliable Kansei ratings across thousands of objects is slow and expensive~\cite{hsu2000semantic}. Prior work on VLM-based aesthetic evaluation has also focused primarily on photographs, where color, texture, lighting, and scene context influence judgments~\cite{datta2006studying,kong2016photo,ke2023vila}. Human agreement remains the long-term validation target, but a simpler question can be asked first: \textit{do independently developed VLMs produce consistent rankings of affective attributes?}

We argue that cross-model convergence provides a useful first-stage consistency audit for semantic controls. If only one model ranks a set of bottles as ``elegant,'' there is limited evidence that the descriptor produces a reproducible signal across models. Consistency across heterogeneous models provides evidence that an affective descriptor induces similar rankings, while the models differ in architecture, training objectives, and data sources. This perspective is inspired by the logic of convergent validity, where convergence across multiple measurements provides stronger evidence than observations from a single source~\cite{campbell1959convergent}. It is also consistent with findings that independently trained neural networks can exhibit measurable representational similarity~\cite{kornblith2019similarity}, suggesting that diverse models may converge toward shared latent representations~\cite{huh2024platonic}. Cross-model convergence does not establish correspondence with human perception, nor does it prove independence from shared training data. Instead, we treat it as an inexpensive offline estimate of whether an affective semantic control exhibits sufficiently stable behavior across models to warrant consideration in an interactive system.

To study this question, we evaluate 6 state-of-the-art VLMs---CLIP, OpenCLIP, SigLIP2, ALIGN, FLAVA, and Qwen3-VL-Embedding (Table~\ref{tab:encoders})---using multi-view renderings of untextured 3D objects from 10 categories in ShapeNetCore, a large-scale repository of 3D models~\cite{chang2015shapenet}. Geometric adjective pairs (e.g., \textit{tall--short}) serve as positive controls, while unrelated adjective pairs (e.g., \textit{loud--quiet}) establish an empirical null. Our results show that affective dimensions exhibit higher cross-model consistency than the null but lower consistency than geometric properties on average. This convergence is strongly category-dependent: some object classes exhibit consistent affective rankings across models, whereas others display little convergence. We further show that convergence depends not simply on the amount of geometric variation within a category, but on whether the semantic direction aligns with the variation structure represented by the models. These findings suggest that cross-model convergence can identify which affective vocabularies are sufficiently reproducible for a given object class before they are presented as controls in a generative design interface.

This work makes three main contributions:
\begin{enumerate}
\item We introduce an offline cross-model consistency audit for affective semantic controls in AI-assisted design interfaces. The audit is calibrated using geometric positive controls and unrelated adjective pairs that establish an empirical null.
\item We characterize when affective semantic controls exhibit reproducible signals across models, showing that convergence depends primarily on the relationship between category-specific variation and the evaluated semantic direction.
\item We demonstrate how the audit can support the design of generative design interfaces and illustrate the workflow with a UI prototype interface that exposes convergence information alongside semantic controls.
\end{enumerate}

\section{Related work}

Our work sits between several related literatures: affective evaluation with vision-language models, Kansei engineering, semantic directions in embedding spaces, interactive intelligent interfaces, and model agreement as an audit criterion for semantic stability. We review each area and position the present study within these lines of work.

\subsection{Interactive Control Through Semantic Attributes}

Generative design systems increasingly explore semantic attributes as interactive mechanisms for steering generation, allowing designers to refine outputs through concepts such as ``more elegant'' or ``less bulky''. This builds on mixed-initiative human-computer interaction (HCI) principles, which emphasize interpretable controls that allow users to guide intelligent systems while remaining in control~\cite{horvitz1999mixedinitiative,amershi2019guidelines}. Early systems explored relative attribute search~\cite{parikh2011relative,kovashka2015whittlesearch} and semantic deformation handles for 3D models~\cite{chaudhuri2013attribit,yumer2015semantic}. More recent approaches expose semantic sliders in learned latent spaces~\cite{harkonen2020ganspace,dang2021ganslider} and use language-guided optimization for 3D generation~\cite{wong2024promptevolution,liu2023dalle3d}. Semantic controls have also been used to structure exploratory search over generative model outputs, allowing users to steer sampling toward selected semantic facets~\cite{liu2024sample}. Learned 3D latent representations have further been incorporated into interactive design systems to support design exploration and decision-making~\cite{saha2022interactivevehicle}.

These approaches generally assume that semantic controls provide reliable signals for interaction. However, their behavior can vary with both the underlying model and object category. This motivates our offline audit of candidate controls before they are exposed to users. Reviews of interactive machine learning interfaces highlight interface design as central to how users interact with and provide feedback to machine-learning systems~\cite{dudley2018review}. More broadly, research on trust in intelligent systems suggests that information about system reliability can support more appropriately calibrated reliance~\cite{lee2004trust}.

\subsection{Kansei Engineering and Product Form Semantics}

Kansei engineering studies how product forms influence human interpretation and affective impressions \cite{nagamachi1995kansei}. A common approach collects affective descriptors and human evaluations to relate perceived impressions to product forms. Such studies span vehicles, furniture, lighting, and household objects \cite{kansei_car_yogasara2017,
kansei_car_syedmohamed2014,kansei_chair_zhou2023,kansei_chair_cai2025,
kansei_sofa_lamp_li2023,kansei_clock_shergian2015,
kansei_clock_mamaghani2014,kansei_jar_mele2018,kansei_bottle_luo2012,
kansei_bookshelf_fu2020,kansei_bookshelf_lin2024,kansei_cabinet_song2026}.
However, affective judgments can vary across demographic groups, limiting the generality of static human-labeled datasets \cite{hsu2000semantic}.

Recent work also uses language models to expand or organize Kansei-related vocabularies, with partial overlap with conventional elicitation \cite{alcaidemarzal2025llm}. We build on this approach by starting from category-specific adjective pairs in the Kansei literature and using an LLM to expand them into polarized pairs. Whereas traditional Kansei studies use these scales to collect human judgments, we use them to test whether independently developed VLMs produce consistent rankings of form-related semantics.

\subsection{Affective and Aesthetic Evaluation with Vision-Language Models}

Aesthetic and affective image evaluation has progressed from hand-crafted photographic features \cite{datta2006studying} to deep aesthetic rankers trained on large image collections \cite{kong2016photo} and models fine-tuned on user commentary \cite{ke2023vila}. Closest to our probe design, CLIP-IQA
\cite{wang2023clip} evaluates abstract image attributes using antonym prompt pairs within a single VLM space.

Our setting differs in two important respects. First, these methods primarily evaluate natural photographs, where color, material, lighting, and scene context influence judgments. Our untextured multi-view 3D renderings instead isolate geometric form, which is particularly relevant to early-stage 3D design. Second, existing approaches generally evaluate a single model as an aesthetic predictor or scoring function, often against human-labeled data. We
instead ask whether affective controls produce reproducible rankings across independently developed VLMs.

\subsection{Semantic Directions in Embedding Spaces}

Modeling concepts as directions in embedding spaces is well established. Word embeddings showed that vector offsets can encode relational structure \cite{mikolov2013linguistic}, while word pairs have been used to isolate interpretable semantic axes \cite{bolukbasi2016man}. Semantic projection extends this idea to continuous attributes by projecting representations onto antonym-defined directions \cite{grand2022semantic}. In VLM spaces, text-vector
differences have similarly been used for image manipulation
\cite{patashnik2021styleclip} and zero-shot aesthetic scoring
\cite{wang2023clip}.

Our probe construction builds directly on this semantic projection framework (Eq.~\ref{eq:direction}). We apply it to test whether candidate affective directions remain sufficiently stable across models to function as interactive controls for a given object category.

\subsection{Cross-Model Agreement as an Audit Criterion}

Agreement across multiple measurements has long been used as evidence in measurement theory \cite{campbell1959convergent}. Related ideas appear in machine learning: deep ensembles use consensus to estimate uncertainty \cite{lakshminarayanan2017simple}, representational similarity metrics compare models trained with different objectives \cite{kornblith2019similarity}, and recent work suggests that large vision models may converge toward shared latent structures \cite{huh2024platonic}.

We build on this rationale but use cross-model agreement for an
interaction-oriented purpose. Rather than asking whether models are generally similar, we ask whether their agreement is sufficient to justify exposing a semantic direction as an interface control. Calibrating affective agreement against geometric positive controls and an empirical null turns model convergence into an offline audit for intelligent design interfaces.

\section{Methodology}
\label{sec:methodology}

\subsection{Overview}
\label{sec:overview}

We conduct a representational audit to examine whether pretrained vision-language models (VLMs) consistently encode affective (Kansei) semantic structure associated with three-dimensional form. All models are evaluated in their pretrained state, without fine-tuning or additional training. 

The methodology consists of four stages, the last of which comprises two complementary evaluation measures (Sections~\ref{sec:convergence} and~\ref{sec:alignment}):

\begin{itemize}
    \item \textbf{Stimulus Rendering:} Objects from 10 \textit{ShapeNetCore} categories are rendered as untextured, uniform grey shapes from 8 fixed viewpoints. By removing surface appearance cues such as color, material, and texture, the rendered stimuli isolate three-dimensional form while preserving geometric information across multiple views.

    \item \textbf{Representation Extraction:} Each rendered view is encoded using 6 pretrained vision-language models. The resulting view-level embeddings are averaged and unit-normalized to produce a single object-level representation that captures information consistent across viewpoints.

    \item \textbf{Directional Projection:} For each affective concept, represented by a bipolar adjective pair (e.g., \textit{modern--traditional}), a semantic direction is constructed by subtracting the corresponding text embeddings within the model's shared vision-language space. Each object representation is then projected onto this direction, producing a continuous score that reflects its position along the semantic axis.

    \item \textbf{Evaluation:} The resulting semantic projections are assessed through \textit{cross-model convergence} (Section~\ref{sec:convergence}), which measures whether different encoders assign consistent relative rankings to objects along the same semantic axis. A further analysis (Section~\ref{sec:alignment}) then asks why convergence varies, testing whether the most convergent directions align with the dominant dimensions of within-category shape variation.
\end{itemize}

\subsection{Stimuli and Rendering}
\label{sec:stimuli}
The target objects are drawn from the \textit{ShapeNetCore} database~\cite{chang2015shapenet} across 10 categories: \textit{chair, table, lamp, sofa, cabinet, bookshelf, bottle, jar, clock,} and \textit{car}. For each category we randomly sample up to $500$ unique objects with a fixed seed, using the full population where it is smaller. In fact, 8 categories reach the $500$ cap; bottle and bookshelf are slightly smaller ($498$ and $452$), giving $4{,}950$ objects in total.

To capture each object's geometry, we render it from 8 viewpoints:
\begin{itemize}
    \item 4 views at $15^\circ$ elevation with azimuths of $0^\circ$, $90^\circ$, $180^\circ$, and $270^\circ$.
    \item 4 views at $35^\circ$ elevation with azimuths of $45^\circ$, $135^\circ$, $225^\circ$, and $315^\circ$.
\end{itemize}
All renders are generated at $512 \times 512$ pixel resolution on a flat white background.

We render all objects using a uniform matte grey material without textures, allowing us to let VLMs evaluate from shape alone. It also avoids confounding associations, such as linking ``luxurious'' to glossy dark surfaces or ``cheap'' to colorful plastic-like appearances, which could mask the contribution of geometry. This also fits the stage of the design process that the audit is intended to support.

Before rendering, each 3D mesh is centered and scaled uniformly by its maximum vertex norm. Uniform scaling preserves the aspect ratios and proportions of the objects.

The overall representation of object $i$ is defined as the mean of its 8 view embeddings, renormalized to unit length. Formally, for an object $i$ with per-view image embeddings $\mathbf{v}_{i,k}$ ($k = 1, \dots, 8$):
\begin{equation}
\bar{\mathbf{v}}_i = \frac{1}{8}\sum_{k=1}^{8} \mathbf{v}_{i,k}, \qquad
\mathbf{e}_i = \frac{\bar{\mathbf{v}}_i}{\lVert \bar{\mathbf{v}}_i \rVert}
\label{eq:normalize}
\end{equation}
We average the view embeddings so that the object representation depends on shape shared across viewpoints rather than on any single canonical view. Normalizing the final vector to unit length ensures that the projection scores (Section~\ref{sec:directions}) measure angular alignment with a semantic vector rather than the absolute magnitude of the image embedding.

\subsection{Encoders}
\label{sec:encoders}
Our audit evaluates 6 vision-language encoders, summarized in Table~\ref{tab:encoders}. All models are used with publicly released weights and evaluated without fine-tuning.

\begin{table*}[t]
\centering
\caption{Overview of evaluated vision-language encoders. The year in parentheses is that of the evaluated checkpoint's public release, which does not always match the year of the cited paper: the OpenCLIP weights predate the scaling-laws paper, and the ALIGN checkpoint is Kakao Brain's open reimplementation trained on COYO-700M rather than the unreleased model of Jia et al.}
\label{tab:encoders}
\small
\setlength{\tabcolsep}{5pt}
\begin{tabularx}{\textwidth}{@{}llclX@{}}
\toprule
\textbf{Name} & \textbf{Model Identifier} & \textbf{Dim.} & \textbf{Architecture} & \textbf{Objective \& Corpus} \\
\midrule
CLIP (2022) & \ident{openai/clip-vit-large-patch14} & 768 & ViT-L/14 & Contrastive InfoNCE~\cite{radford2021clip} \\
OpenCLIP (2022) & \ident{laion/CLIP-ViT-H-14-laion2B-s32B-b79K} & 1024 & ViT-H/14 & Contrastive, LAION-2B~\cite{cherti2023reproducible} \\
SigLIP2 (2025) & \ident{google/siglip2-so400m-patch14-384} & 1152 & SoViT-400m/14 & Pairwise sigmoid loss~\cite{tschannen2025siglip2} \\
ALIGN (2023) & kakaobrain/align-base & 640 & EfficientNet + BERT &
Noisy web supervision~\cite{jia2021scaling}, COYO-700M~\cite{byeon2022coyo} \\
FLAVA (2022) & \ident{facebook/flava-full} & 768 & ViT + multimodal fusion & Multimodal masked objective~\cite{singh2022flava} \\
Qwen3-VL-Emb. (2026) & \ident{Qwen/Qwen3-VL-Embedding-2B} & 2048 & LLM backbone & Retrieval/distillation~\cite{li2026qwen3vlembedding} \\
\bottomrule
\end{tabularx}
\end{table*}

Our framework relies on structural diversity among encoders. If the models were incremental variants of a single structure, agreement could be attributed to shared training biases or architectural constraints. We therefore select models that differ across several dimensions:
\begin{itemize}
    \item \textbf{Training Objective:} CLIP, OpenCLIP, and ALIGN employ contrastive InfoNCE objectives. SigLIP2 replaces the softmax normalization over the global batch with a pairwise sigmoid loss~\cite{tschannen2025siglip2}. FLAVA relies on multimodal masked modeling~\cite{singh2022flava}. Qwen3-VL-Embedding uses a text-retrieval objective optimized via contrastive learning and reranker distillation~\cite{li2026qwen3vlembedding}.
    \item \textbf{Architectural Inductive Bias:} 5 models use Vision Transformers (ViT) of varying scales and patch sizes, while ALIGN uses a convolutional EfficientNet~\cite{jia2021scaling} paired with a BERT-style text tower.
    \item \textbf{Training Provenance and Scale:} The models originate from different organizations (OpenAI, LAION, Google, Kakao Brain, Meta, Alibaba) and were trained on distinct web-scale datasets under different filtering regimes, with embedding dimensionalities ranging from $640$ to $2048$.
\end{itemize}
For inference, we follow each model's official documentation (e.g., CLIP-family pooling and Qwen3-VL-Embedding's last-token pooling).

\subsection{Probe Design}
\label{sec:probes}
Our probes consist of bipolar word pairs organized into 3 tiers.

\paragraph{Tier 1 --- Geometric (7 pairs; 70 instances).}
We select visually verifiable physical descriptors that serve as positive controls: \textit{tall--short, boxy--curvy, simple--complex, elongated--compact, thin--thick, wide--narrow,} and \textit{angular--rounded}. If our pipeline cannot reliably capture basic shape attributes, negative results on more abstract concepts would be uninterpretable. These 7 axes name roughly 3 underlying concepts, namely extent, curvature, and complexity, so we treat their agreement as a robustness check.

\paragraph{Tier 2 --- Kansei (32 pairs; 78 instances).}
This is our primary tier. It contains stylistic and affective descriptors,
including 3 ``shared-core'' pairs evaluated across all 10 categories (\textit{modern--traditional, elegant--messy, luxurious--cheap}) and
category-specific pairs (e.g., \textit{plush--rigid} for sofas,
\textit{delicate--robust} for lamps). Category-specific pairs were drawn from
prior Kansei engineering studies of the corresponding product classes or, where no exact-category study was available, closely related product classes:
cars~\cite{kansei_car_yogasara2017, kansei_car_syedmohamed2014, kansei_car_jindo1997},
chairs~\cite{kansei_chair_zhou2023, kansei_chair_cai2025},
sofas~\cite{kansei_sofa_lamp_li2023},
clocks~\cite{kansei_clock_shergian2015, kansei_clock_mamaghani2014},
bookshelves~\cite{kansei_bookshelf_fu2020, kansei_bookshelf_lin2024},
cabinets~\cite{kansei_cabinet_song2026},
and bottles~\cite{kansei_bottle_luo2012, kansei_jar_mele2018}. We then follow a hybrid procedure, taking these literature-sourced terms as seeds and applying an LLM-assisted method~\cite{alcaidemarzal2025llm} both to extend the corpus with
additional descriptors for each category and, for categories without a dedicated study, to generate the descriptor set directly; in all cases the method also assigns a polar opposite to every term. This combines the domain grounding of published Kansei studies with the coverage of LLM-generated vocabularies, which have been shown to produce Kansei semantic spaces equivalent
to those built with conventional techniques~\cite{alcaidemarzal2025llm}. Pairs
whose poles differ only by a negating affix (e.g., \textit{comfortable--uncomfortable})
were excluded, since their text embeddings are near-identical.

\paragraph{Tier 3 --- Irrelevant (20 pairs; 200 instances).}
This tier comprises sensory and abstract antonyms, such as \textit{loud--quiet, sweet--sour,} and \textit{hot--cold}, that have no plausible physical connection to static, grey 3D shapes. It serves as an empirical null for calibrating our metrics and establishing the baseline of spurious agreement.

\subsection{Directions and Projection}
\label{sec:directions}
For encoder $e$ and word pair $p = (\text{word}, \text{antonym})$, let $\mathbf{t}_w^{(e)}$ denote the text embedding of adjective $w$ under encoder $e$, obtained by encoding the bare adjective with that model's text encoder. We extract the embeddings of both poles and compute their normalized difference direction:
\begin{equation}
\mathbf{d}_p^{(e)} = \frac{\mathbf{t}_{\text{word}}^{(e)} - \mathbf{t}_{\text{antonym}}^{(e)}}{\lVert \mathbf{t}_{\text{word}}^{(e)} - \mathbf{t}_{\text{antonym}}^{(e)} \rVert}
\label{eq:direction}
\end{equation}
where $\mathbf{d}_p^{(e)}$ is a unit vector pointing from the negative pole to the positive pole. Using vector differences to represent semantic relations builds on established approaches in word embedding research~\cite{mikolov2013linguistic, bolukbasi2016man}. This approach is closely related to semantic projection, in which semantic axes defined by opposing word pairs are used to recover continuous properties from word embeddings~\cite{grand2022semantic}. Applying it in CLIP space rests on the assumption that text and image representations move in roughly parallel directions for the same concept, as used in text-driven manipulation methods~\cite{patashnik2021styleclip}. Antonym prompt pairs have likewise been used to evaluate visual and perceptual image traits~\cite{wang2023clip}.

The score of object $i$ on axis $p$ within encoder $e$ is:
\begin{equation}
s_{p,i}^{(e)} = \langle \mathbf{e}_i^{(e)}, \mathbf{d}_p^{(e)} \rangle
\label{eq:projection}
\end{equation}
and $\mathbf{s}_p^{(e)} \in \mathbb{R}^N$ denotes the vector of projection scores for all $N$ objects in a category.

This design addresses two requirements:
\begin{itemize}
    \item \textbf{Canceling shared subspaces:} Constructing axes as vector differences is essential because individual word vectors are dominated by a shared anisotropic component present across almost all embeddings. Subtracting the antonym cancels out this common offset.
    \item \textbf{Preserving internal metrics:} Because directions are calculated independently within each model's native coordinate system, the absolute scores $s_{p,i}^{(e)}$ are not directly comparable across different models (e.g., a score of $0.2$ in CLIP does not mean the same as $0.2$ in Qwen3-VL). Consequently, our evaluation metrics rely on rank orderings rather than absolute values.
\end{itemize}

\subsection{Cross-Model Convergence}
\label{sec:convergence}
Our proposed metric, cross-model convergence, measures how consistently the 6 encoders order objects along a given axis. For each encoder pair, we compute the Spearman rank correlation of their object scores and average over all $\binom{6}{2}=15$ unique pairs:
\begin{equation}
\bar{\rho}_p = \binom{E}{2}^{-1} \sum_{e < e'} \rho_S\left( \mathbf{s}_p^{(e)}, \mathbf{s}_p^{(e')} \right)
\label{eq:convergence}
\end{equation}
where $E = 6$ and $\rho_S$ is Spearman's rank correlation. 

This formulation aligns with classical convergent validity~\cite{campbell1959convergent}: a construct is considered validly measured when independent assessment methods produce consistent results.

\paragraph{Statistical Evaluation.}
To compare two tiers, we use a one-sided Wilcoxon rank-sum (Mann--Whitney $U$) test, which accommodates unequal sample sizes, non-normal distributions, and our directional hypotheses (e.g., Kansei axes should show higher convergence than irrelevant axes). We report the common-language (CL) effect size:
\begin{equation}
\text{CL} = \frac{U}{n_1 n_2}
\label{eq:cl_effect}
\end{equation}
which represents the probability that a randomly selected axis from the target tier has higher convergence than one from the comparison tier, and we report the two-sample Kolmogorov--Smirnov (KS) statistic to characterize general distributional differences.

We run many tier comparisons across axes and categories, and the axes are not fully independent, since categories share the null set and the core pairs, and axes within a category share object renderings. We therefore do not treat individual $p$-values as the basis for our claims. We report common-language effect sizes, which depend only on rank overlap, and we quantify the sampling variability of per-category convergence with a bootstrap: for each category we resample its objects with replacement over $1{,}000$ replicates, recompute convergence on each resample, and report the $2.5$ and $97.5$ percentiles as a $95\%$ confidence interval. Differences we describe as category-dependent are those whose intervals do not overlap.

\subsection{Variation-Subspace Alignment}
\label{sec:alignment}
Convergence establishes \emph{whether} an axis produces consistent rankings, but not \emph{why} some axes converge and others do not. We hypothesize that an axis converges when its direction lies within the subspace along which a category's shapes actually vary, and test it.

For each category and encoder $e$, we compute the top $K = 20$ principal components $\{\mathbf{u}_1^{(e)}, \dots, \mathbf{u}_K^{(e)}\}$ of the mean-centered object embeddings, with variance ratios $\lambda_k^{(e)}$. Each $\mathbf{u}^{(e)}_k$ is a unit vector in encoder~$e$'s embedding space
along the $k$-th direction of greatest variation among that category's objects. For a unit direction $\mathbf{d}_p^{(e)}$, we measure its alignment with this variation subspace as a variance-weighted projection:
\begin{equation}
A_p^{(e)} = \sum_{k=1}^{K} \lambda_k^{(e)} \,
\bigl\langle \mathbf{d}_p^{(e)}, \mathbf{u}_k^{(e)} \bigr\rangle^2 .
\label{eq:alignment}
\end{equation}
$A_p^{(e)}$ is large when the direction concentrates along the dominant modes of the category and small when it points into low-variance directions. We average $A_p^{(e)}$ across encoders and correlate it with convergence $\bar{\rho}_p$ using Spearman's rank correlation, separately per tier. The principal components capture the directions of greatest variation among the object embeddings; they do not necessarily correspond to named or visually interpretable geometric axes. We therefore read $A_p^{(e)}$ as a measure of how much of a direction lies within the space the models use to separate objects, and we report its relationship to convergence as a correlation rather than a causal mechanism. As a baseline, we repeat both measurements (convergence and alignment) on random unit directions drawn uniformly in each encoder's embedding space, matched in number to the affective axes. These carry no semantic content and establish the values expected when a direction has no relation to either the category's variation or the models' shared structure.

\subsection{View Reliability}
\label{sec:reliability}
Because each object representation is averaged over only 8 views (Eq.~1), we assess whether the resulting rankings are stable across subsets of viewpoints. For each category and encoder~$e$, we split the eight views into two disjoint halves of 4 views, construct an object representation from each half, and project both representations onto every direction. Let $r^{(e)}_{\text{hh}}$ denote the Spearman correlation between the two resulting score vectors across objects. Since each half rests on only 4 views, we apply the Spearman--Brown correction to estimate the reliability of
the full eight-view representation,

\begin{equation}
r^{(e)}_{\text{SB}} = \frac{2\,r^{(e)}_{\text{hh}}}{1 + r^{(e)}_{\text{hh}}},
\end{equation}

and report the mean of $r^{(e)}_{\text{SB}}$ per category. To account for measurement unreliability when assessing convergence, we disattenuate each pairwise correlation by the reliabilities of the two encoders involved,

\begin{equation}
\tilde{\rho}^{(e,f)}_p =
\frac{\rho^{(e,f)}_p}{\sqrt{r^{(e)}_{\text{SB}}\, r^{(f)}_{\text{SB}}}},
\end{equation}

where $\rho^{(e,f)}_p$ is the raw agreement between encoders $e$ and $f$ on direction~$p$. A convergence value that remains after this correction is therefore less likely to be explained by view-sampling noise alone.

\subsection{Object Specificity}
\label{sec:specificity}
A category-specific affective axis (e.g., \textit{plush--rigid} for sofas) is authored for one object class. If the affective signal were a generic property of the text embedding rather than tied to the object it describes, such an axis would converge equally well when applied to any category. To test this, we apply each category-specific axis both to its own category (\emph{own}) and to every other category (\emph{foreign}), and compare the mean convergence of the two groups. We exclude the 3 shared-core pairs from this analysis, since they are used for all categories and would otherwise inflate the own-category mean. Higher own-category than foreign-category convergence would indicate that the affective signal is specific to the object class the axis was written for.

\subsection{Prompting Sensitivity}
\label{sec:prompt_sensitivity}
Our probes are bare adjectives paired with their antonyms. Two alternative constructions are plausible: first, we replace the bipolar difference direction (Eq.~\ref{eq:direction}) with a single-pole word embedding, projecting objects onto the positive-pole vector alone. Second, we replace the bare adjective pair with sentence templates of the form ``a \{adjective\} \{category\},'' embedding the full phrase rather than just the word. In each case we recompute convergence for all 3 tiers.

\subsection{Reproducibility}
All models are used with publicly released weights, and the complete probe vocabulary is given in Appendix~\ref{app:vocab}. The
rendering pipeline, embedding extraction, analysis code, and the
precomputed audit values will be released publicly upon publication. 
\section{Results}
\label{sec:results}

Throughout, convergence is the mean pairwise Spearman correlation $\bar\rho_p$ across the $\binom{6}{2}=15$ encoder pairs (Eq.~\ref{eq:convergence}). Tier comparisons use one-sided Mann--Whitney common-language (CL) effect sizes (Eq.~\ref{eq:cl_effect}). The audit covers $348$ (category, axis) instances: $70$ geometric, $78$ affective, and $200$ irrelevant.

\begin{figure}[t]
  \centering
  \includegraphics[width=0.5\textwidth]
{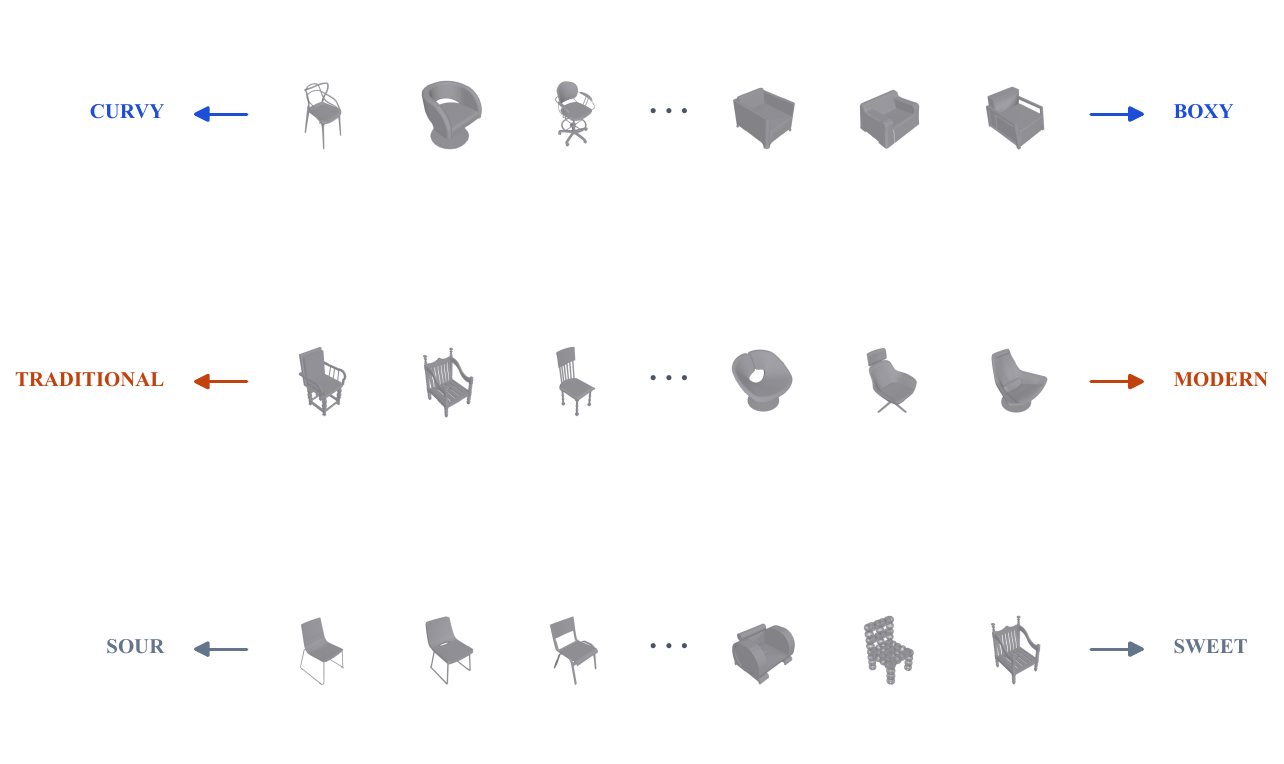}
  \caption{Objects along one axis per tier, for chairs under a
  single encoder (Qwen3-VL-Embedding). Each row shows the 3
  lowest-scoring and 3 highest-scoring objects on that axis.}
  \Description{Three horizontal rows of rendered grey chair models, one row per tier. Each row shows the three lowest-scoring chairs on the left, and the three highest-scoring chairs on the right, with the negative pole labelled at the left end and the positive pole at the right end. The rows are curvy to boxy, traditional to modern, and sour to sweet.}
  \label{fig:extremes}
\end{figure}

\subsection{Convergence across tiers}
\label{sec:res-tiers}

\begin{figure}[t]
  \centering
  \includegraphics[width=0.4\textwidth]{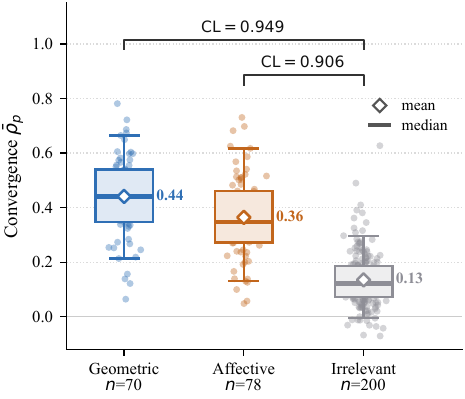}
  \caption{Convergence by tier. Points are individual (category, axis)
  instances; boxes span the interquartile range with whiskers at the 5th and
  95th percentiles}
  \Description{A box plot comparing cross-model convergence across three tiers, with individual data points overlaid. Geometric axes have the highest mean at 0.44, affective axes fall in the middle at 0.36, and irrelevant axes are lowest at 0.13. Brackets above the boxes give common-language effect sizes: 0.949 comparing geometric to irrelevant, and 0.906 comparing affective to irrelevant.}
  \label{fig:tiers}
\end{figure}

The tiers order as expected (Figure~\ref{fig:tiers}): geometric pairs converge most ($\bar\rho = 0.441$), the irrelevant null least ($0.135$), and affective ones fall between ($0.364$). A randomly chosen affective axis out-converges a randomly chosen irrelevant one about 9 times in 10 ($\mathrm{CL}=0.906$, KS $D=0.70$); geometric controls beat the null more strongly ($\mathrm{CL}=0.949$, KS $D=0.81$).

The gap between geometric and affective axes is real but small
($\mathrm{CL}=0.658$, KS $D=0.33$), and the weakest geometric axis
(\emph{wide--narrow}, $\bar\rho=0.36$) sits at the Kansei mean. Together, these results show that the encoders agree on affective orderings almost as consistently as on geometric ones, even though the stimuli carry no colour, material, or texture. 

Agreement is not driven by encoder redundancy. OpenCLIP reproduces the CLIP objective and architecture, so this pair could inflate the result. It does not: among affective axes CLIP--OpenCLIP agree at $\rho = 0.34$, below the mean pairwise value, while the strongest agreement is shared by SigLIP2--Qwen3-VL-Embedding and SigLIP2--ALIGN ($\rho = 0.47$), pairs that share neither architecture nor training goal. Dropping OpenCLIP raises the affective mean only from $0.365$ to $0.379$ (full matrix in Appendix~\ref{app:matrix}).

\subsection{Convergence depends on both the axis and the category}
\label{sec:res-category}

\begin{figure*}[t]
\centering
\includegraphics[width=0.8\textwidth]{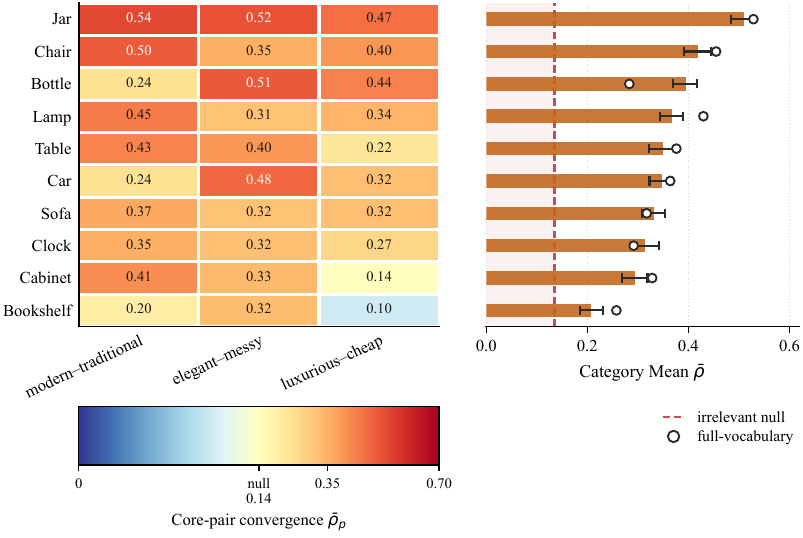}
\caption{Shared-core affective convergence by category. \textbf{Left:}
Cross-model convergence $\bar\rho_p$ for each category on the 3 shared-core axes (\textit{modern--traditional}, \textit{elegant--messy}, \textit{luxurious--cheap}), coloured relative to the irrelevant null ($0.14$): warm cells lie above chance, cool cells at or below it. \textbf{Right:} Mean convergence over the 3 shared-core axes per category (bars), ordered high to low, with $95\%$ confidence intervals over objects; the shaded region marks convergence at or below the null. Hollow markers show each category's full-vocabulary mean. Categories are ordered by shared-core convergence; bottles rank third here but ninth on their full vocabulary.}
\Description{A two-panel figure. The left panel is a heatmap of ten object categories by the three shared-core affective axes, with cells coloured relative to the irrelevant null of 0.14; warm cells indicate convergence above chance and cool cells at or below it. Jars have the highest values and bookshelves the lowest. 
The right panel is a horizontal bar chart of each category's mean
convergence over the three shared-core axes, ordered from high to low with ninety-five percent confidence intervals, a dashed line marking the irrelevant null, and hollow markers showing each category's full-vocabulary mean.}
\label{fig:category}
\end{figure*}

Affective convergence varies widely across object classes
(Figure~\ref{fig:category}). We report two rankings. The full-vocabulary ranking averages each category over its own Kansei axes, which differ by category and are therefore not comparable across classes. The shared-core ranking uses only the 3 axes evaluated everywhere (\emph{modern--traditional}, \emph{elegant--messy}, \emph{luxurious--cheap}).

On the shared core, convergence ranges from $\bar\rho=0.21$ (bookshelf) to $0.51$ (jar). The full-vocabulary ranking spans a similar range ($0.26$--$0.53$) and correlates with the shared pairs ($\rho=0.71$), but individual categories move: bottles rank third on the shared core and ninth on their full vocabulary.

The vocabulary matters as much as the object class. \emph{Minimalist--ornate}
($\bar\rho=0.53$) and \emph{elegant--messy} ($0.39$) hold up, while
\emph{luxurious--cheap} drops to $0.30$, \emph{masculine--feminine} to $0.24$, and \emph{natural--artificial} ($0.14$) reaches the null. The
geometric-over-affective ordering also shifts: geometric axes dominate for bookshelves, clocks, and cabinets, the tiers are comparable for chairs and lamps, and affective axes win for jars ($0.53$ vs.\ $0.47$) and marginally cars ($0.36$ vs.\ $0.28$). Consistency is then a property of the (axis, category) pair, not of either alone.

Each object representation averages 8 views, so we check whether these different renderings reflect stable rankings. Split-half reliability (Spearman--Brown corrected) averages $0.76$, from $0.61$ (bookshelf) to $0.89$ (jar). Reliability correlates with raw convergence ($\rho=0.47$), so part of the category effect could be measurement quality. Disattenuating each encoder pair by its split-half reliability breaks that link: corrected convergence is uncorrelated with reliability ($\rho=0.10$), while the category spread remains ($0.32$--$0.58$).

Bookshelves show the difference between low agreement and unreliable measurement. They have the lowest shared-core convergence ($0.21$) and the lowest reliability ($0.61$), but that reliability sits above the noise floor and their cross-half rankings are stable. Disattenuating each pairwise correlation raises their cross-model convergence value, still the lowest in the set: correction lifts the whole tier without changing bookshelves' position relative to other
categories.

\begin{table}[t]
\centering
\caption{Convergence before and after disattenuation by split-half reliability.}
\label{tab:disattenuation}
\small
\begin{tabular}{lcc}
\toprule
Tier & Raw $\bar\rho$ & Corrected $\bar\rho$ \\
\midrule
Geometric   & 0.441 & 0.548 \\ 
Affective   & 0.364 & 0.464 \\
Irrelevant  & 0.135 & 0.184 \\
\bottomrule
\end{tabular}
\end{table}

\subsection{Alignment with shape variation}
\label{sec:res-mechanism}

We test whether an axis converges when a category's shape variation lies along it, using the alignment $A_p$ between the direction and the top-$20$ principal components of the object embeddings (Eq.~\ref{eq:alignment}), averaged across encoders. 

\begin{figure}[H]
\centering
\includegraphics[width=0.4\textwidth]{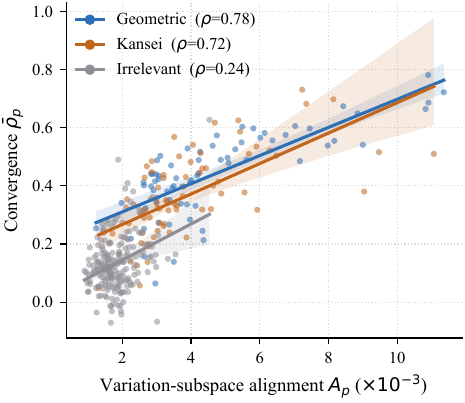}
\caption{Variation-subspace alignment. Each point is a (category, axis) instance; lines are least-squares fits with $95\%$ confidence bands. Geometric and affective slopes are similar; the irrelevant tier lies below both.}
\Description{A scatter plot of cross-model convergence against
variation-subspace alignment, with one point per category-axis instance, coloured by tier, and least-squares fit lines with ninety-five percent confidence bands. All three tiers show a positive relationship: geometric axes have a rank correlation of 0.78, affective axes 0.72, and irrelevant axes 0.24. The geometric and affective fit lines have similar slopes and lie above the
irrelevant line.}
\label{fig:alignment}
\end{figure}

Alignment predicts convergence (Figure~\ref{fig:alignment}): $\rho=+0.78$ for geometric axes, $+0.72$ for affective axes, and $+0.24$ for the irrelevant tier ($p<0.001$ throughout). The similar slopes suggest one statistical relationship that geometry satisfies more often. Kansei axes align with real object variation far more than random directions do (random: ${\approx}\,8\times10^{-4}$; Kansei: up to $1.1\times10^{-2}$). Only the weakest affective axes fall to the random level, and those are indeed the ones that do not produce agreement.
Random directions show neither an alignment--convergence relationship
($\rho = -0.04$, $p = 0.56$) nor cross-model convergence ($\bar\rho = -0.003$).

Alignment and projection variance are strongly correlated ($\rho=+0.74$ for variance alone), so Equation~\ref{eq:alignment} might simply restate variance: a direction in a low-variance subspace produces little score spread, its ranking is noise-dominated, and convergence is attenuated. Even after removing the effect of score spread and correcting each axis for its own measurement noise, alignment still predicts convergence (Kansei $\rho = 0.47$, $p < 0.001$; geometric $0.88$; irrelevant $0.15$). The relation also holds within categories: across all 3 tiers per class, $\rho$ runs from $0.66$ (car) to $0.87$ (lamp), positive in all ten; restricted to affective axes alone it stays positive in every category (median $\rho = 0.79$). 

For clocks, bookshelves, and cabinets, geometric axes converge far more strongly than affective ones ($0.54$, $0.49$, $0.45$ versus $0.29$, $0.26$, $0.33$). Their shapes vary enough for the models to agree on geometric orderings, but that variation does not lie along the affective directions. What matters is not whether objects differ, but whether they differ along the direction the semantic dimension names.

One possible explanation could be that the convergent Kansei axes simply capture the same information as our geometric pairs. We test this by computing cosine similarity between axes: within each encoder, the affective axes do not overlap with each other, and they are also independent enough from the 7 geometric axes. The strongest relationship is between the pairs \emph{simple--complex} and \emph{minimalist--ornate} (mean $|\cos| = 0.24$, maximum $0.42$). Even in this case, they share less than one-fifth of their variance. Overall, none of the affective axes aligns with a geometric axis. This shows that the observed convergence captures aspects of Kansei meaning that cannot be explained by the geometric pairs chosen for the experiments.

\begin{table}[H]
\centering
\caption{Direction orthogonality (mean over categories). \emph{Affective within} refers to cosine similarity computed between Kansei axes of the same object category, while \emph{Affective$\times$Geometric} compares Kansei axes with geometric ones. Low $|\cos|$
indicates distinct directions. }
\label{tab:ortho}
\small
\begin{tabular}{lcc}
\toprule
Encoder & Affective within & Affective$\times$Geometric \\
\midrule
ALIGN    & 0.09 & 0.07 \\
CLIP     & 0.11 & 0.08 \\
OpenCLIP & 0.09 & 0.09 \\
SigLIP2  & 0.10 & 0.06 \\
FLAVA    & 0.14 & 0.10 \\
Qwen     & 0.15 & 0.11 \\
\bottomrule
\end{tabular}
\end{table}

\subsection{Robustness to encoders and prompts}
\label{sec:res-controls}

We first test whether convergence among 5 encoders predicts the sixth. For each (category, axis) instance we compute convergence among 5 encoders and, separately, the sixth encoder's mean Spearman correlation with those five. Five-encoder convergence predicts the held-out encoder's agreement for every encoder (Table~\ref{tab:loo}; overall $\rho = 0.67$).

\begin{table}[t]
\centering
\small
\setlength{\abovecaptionskip}{4pt}
\setlength{\belowcaptionskip}{0pt}
\caption{Sixth-encoder prediction.}
\label{tab:loo}
\begin{tabular}{lc}
\toprule
Held-out encoder & $\rho$(5-encoder conv., held-out agreement) \\
\midrule
FLAVA           & 0.60 \\
OpenCLIP        & 0.63 \\
Qwen3-VL-Emb.   & 0.68 \\
ALIGN           & 0.70 \\
CLIP            & 0.70 \\
SigLIP2         & 0.76 \\
\midrule
Overall         & 0.67 \\
\bottomrule
\end{tabular}

\vspace{8pt}

\caption{Effects of different text input construction on tier convergence.}
\label{tab:ablation}
\begin{tabular}{lcccc}
\toprule
Probe form & Geometric & Affective & Null & Aff.$-$Null \\
\midrule
Bipolar           & 0.44 & 0.36 & 0.14 & 0.23 \\
Single-pole       & 0.40 & 0.27 & 0.21 & 0.06 \\
Sentence template & 0.54 & 0.44 & 0.21 & 0.23 \\
\bottomrule
\end{tabular}
\end{table}

\begin{figure*}[t]
\centering
\includegraphics[width=\linewidth]{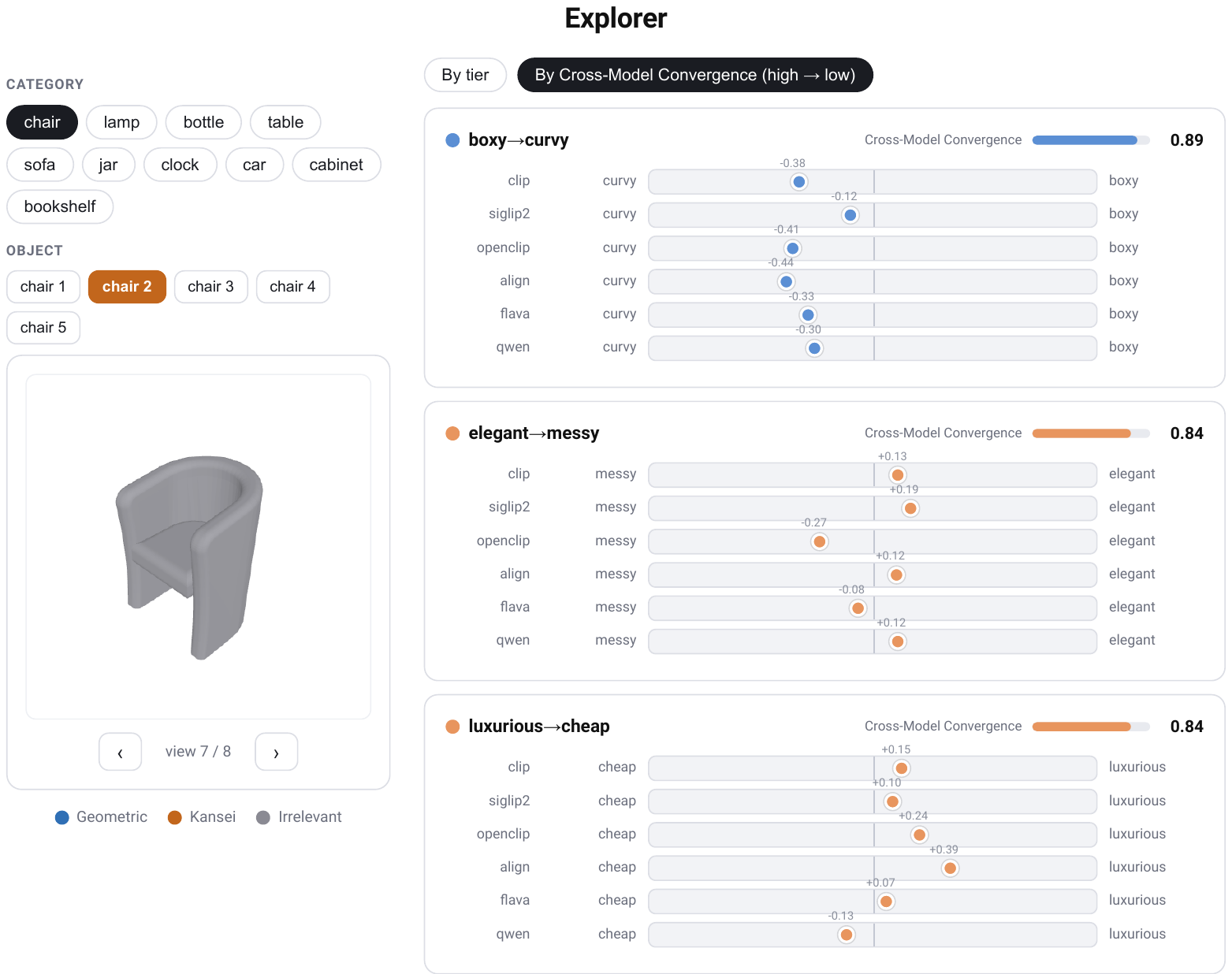}
\caption{A UI prototype built on the precomputed audit. For the selected object, each semantic direction shows where the 6 encoders place that
object, together with an object-level agreement score.}
\Description{A screenshot of the prototype interface. A left sidebar offers buttons for selecting an object category and a specific object, below which a rendered grey chair is shown with controls for stepping through its eight views. The main panel lists semantic axes as horizontal slider tracks, sorted by cross-model convergence. Each axis shows one row per encoder with a marker positioned at that encoder's projection score between the two poles, together with the axis's overall cross-model convergence value. Similarly clustered markers correspond to high convergence.}
\label{fig:demo}
\end{figure*}

If Kansei pairs were a generic text-space effect, an axis written for one category would work equally well on any other. Applying each category-specific axis to its own category and to every other one, own-category convergence exceeds foreign-category convergence for most categories ($0.35$ vs.\ $0.28$; $\mathrm{CL}=0.64$, $p=0.009$, shared-core pairs excluded), reversing for bottles, clocks, and tables, the three weakest categories overall.

Table~\ref{tab:ablation} reports two alternative input text constructions. Replacing the bipolar difference with a single-pole embedding removes the calibration: the null rises to $0.21$ while affective convergence falls to $0.27$, shrinking the affective--null gap from $0.23$ to $0.06$. Individual word vectors are dominated by a shared anisotropic component, and subtracting the antonym removes it.
Sentence templates (``a \{adjective\} \{category\}'') raise all three tiers convergence while preserving the absolute gap ($0.23$), because the shared category noun adds a component related to the object category. This is also more consistent with the text input style used to train these VLMs. We keep bare bipolar pairs, which hold the null near zero.

\section{Discussion}

Our results suggest that cross-model convergence can serve as a practical consistency check for semantic controls in AI-assisted design systems. Descriptors with high cross-model convergence can be treated as more reproducible controls, while those with low convergence can be flagged for further evaluation.

\subsection{From audit to interface}
\label{sec:res-interface}

The audit produces a (category, axis) map example that could be used to select semantic controls for interactive systems (Figure~\ref{fig:category}). The audit shows selectivity at both the vocabulary and category levels (Section \ref{sec:res-category}), reflecting that the same semantic descriptor may exhibit consistent model agreement for one object class but not another. Because the procedure requires neither human labels nor
model retraining, incorporating a newly released encoder requires only
recomputing projections over the existing rendered dataset.

Figure~\ref{fig:demo} illustrates one possible use of such an audit in an interactive system. A user selects an object from a category, and each semantic axis displays where the 6 encoders project that object together with the corresponding per-object agreement score. Similarly clustered encoder scores show higher cross-model convergence values, whereas low-convergence axes reveal larger disagreement. The interface visualizes the precomputed audit
described in Section~\ref{sec:results}.

The purpose of this interface example is to assist user judgment by making model agreement visible during interaction. Moreover, existing semantic controls of such systems often implicitly assume that descriptors such as \emph{elegant} or \emph{luxurious} behave consistently across objects. Our results indicate that this consistency depends strongly on object category. Presenting convergence alongside each semantic control could help users assess whether a control reflects a pattern shared across VLMs or is specific to a particular model. This design is motivated by prior evidence that communicating model uncertainty can encourage more analytical engagement and reduce over-reliance on system output~\cite{prabhudesai2023understanding}. It could also provide evidence about whether a given descriptor elicits consistent responses across VLMs for a particular object category.

The audit supports a concrete admission rule: the irrelevant tier establishes an empirical floor at $\bar{\rho}=0.14$, and bootstrap intervals are computed for each category. A (category, axis) sample can therefore be classified as: expose the control when the lower interval bound exceeds the null; mark it as provisional when the interval overlaps the null; and withhold it when the estimate is at or below the null. In Figure~\ref{fig:category}, this rule implies to expose \emph{luxurious--cheap} for jars ($\bar{\rho}=0.47$) but withhold it for bookshelves ($0.10$) and cabinets ($0.14$). Withholding means that the encoders provide insufficiently reproducible evidence across VLMs to use it as a control. A weak result can also trigger substitution: when a descriptor is near the null, the interface can suggest a better-supported alternative with close semantic meaning. This preserves user control while making model uncertainty explicit.

We see the audit as infrastructure for generative design interfaces rather than as an end-user evaluation. Future systems could use convergence to rank candidate semantic controls, highlight descriptors that perform close to the empirical null, or communicate uncertainty about a semantic edit to users of generative 3D modeling systems. 

\subsection{Practical and Societal Implications}

The proposed audit can be used as an inexpensive preprocessing step when constructing semantic controls for generative design interfaces. This exposes where models disagree, helping designers see whether a descriptor reflects a pattern shared across encoders or the behavior of one model. However, the evaluated VLMs may share training data, linguistic conventions, and representational biases, while the Kansei vocabulary includes descriptors whose meanings can vary across cultures, demographic groups, and design contexts. Low-agreement controls should therefore be flagged as uncertain rather than removed, so that designers' attention can be focused on judging whether the descriptor is appropriate. Before use in culturally sensitive or consequential settings, selected descriptors should be checked with the intended user group.
\section{Limitations}

\paragraph{No human validation.}
The proposed audit measures cross-model consistency rather than agreement with human affective judgements. Consequently, our results should not be interpreted as evidence that the recovered semantic directions correspond to human Kansei perception. Instead, we treat cross-model convergence as a model-based consistency criterion that can assist human judgements. Establishing how convergence relates to it is a direction for future work.

\paragraph{The interface has not been evaluated with users.}
The examples demonstrate that cross-model convergence can be incorporated into an interface to expose model agreement and to support design choices such as flagging or filtering semantic controls. However, we do not evaluate whether these mechanisms improve designer performance, decision quality, calibration, or trust. Determining the practical value of convergence-aware interfaces will require controlled user studies comparing them with conventional semantic-control interfaces.

\section{Conclusion}

We presented a cross-model convergence framework for auditing affective
semantic controls in AI-assisted design. Using 6 independently developed vision-language models, we showed that affective axes exhibit, on average, higher agreement than an empirical null while remaining below geometric controls on average. We further found that cross-model agreement depends on whether the semantic direction captures variation that is actually present among the objects being compared.

Rather than treating cross-model agreement as a substitute for human evaluation, we argue that it provides a useful first-stage consistency assessment for semantic controls. The proposed audit requires no human labels, scales to new object categories and models, and can be recomputed as vision-language models continue to evolve. By identifying which affective descriptors exhibit reproducible cross-model behavior before they are exposed to users, the framework provides a practical foundation for building more transparent and uncertainty-aware intelligent design interfaces.

Future work should validate cross-model convergence against human Kansei judgements, investigate whether the Kansei words can be optimized, explore richer representations that incorporate materials, textures, and contextual factors, and evaluate how presenting semantic reliability information influences designer decision-making in interactive systems.

\section*{GenAI Usage Disclosure}
We used large language models (LLMs) in the preparation of this work in the following ways. First, to support construction of the Kansei probe vocabulary: where a category lacked sufficient affective descriptors from the source literature, we prompted an LLM to suggest candidate adjectives and to propose bipolar antonyms for existing terms; all suggested pairs were reviewed and curated by the authors before use, and the final vocabulary is reported in full. Second, for language editing: we used an LLM to polish phrasing and improve clarity of the manuscript text. Third, we used an LLM-based coding assistant to support figure-generation. All experimental design, data analysis, interpretation of results, and scientific claims are the authors' own, and the authors take full responsibility for the content of the paper.
\bibliographystyle{ACM-Reference-Format}
\begin{acks}
This work was funded by the European Union under Horizon Europe (Grant Agreement No. 101226927). Views and opinions expressed are however those of the author(s) only and do not necessarily reflect those of the European Union or the European Research Executive Agency (REA). Neither the European Union nor the granting authority can be held responsible for them.  
www.genaide.eu
\end{acks}
\bibliography{references}
\appendix
\afterpage{\clearpage}
\section{Encoder Agreement Matrix}
\label{app:matrix}
Table~\ref{tab:matrix} reports the mean pairwise Spearman rank correlation
between every pair of encoders, averaged over all Kansei axes.

\section{Vocabulary}
\label{app:vocab}
The complete probe vocabulary is given in Table~\ref{tab:vocab}.

\begin{table*}[p]
\centering
\caption{Mean pairwise Spearman agreement between encoders across affective axes.}
\Description{Symmetric six-by-six matrix of mean Spearman correlations between encoder pairs.}
\label{tab:matrix}
\small
\setlength{\tabcolsep}{4pt}
\begin{tabular}{lcccccc}
\toprule
& CLIP & OpenCLIP & SigLIP2 & ALIGN & FLAVA & Qwen \\
\midrule
CLIP     & 1.00 & 0.34 & 0.41 & 0.36 & 0.34 & 0.39 \\
OpenCLIP & 0.34 & 1.00 & 0.41 & 0.35 & 0.26 & 0.32 \\
SigLIP2  & 0.41 & 0.41 & 1.00 & 0.47 & 0.31 & 0.47 \\
ALIGN    & 0.36 & 0.35 & 0.47 & 1.00 & 0.31 & 0.38 \\
FLAVA    & 0.34 & 0.26 & 0.31 & 0.31 & 1.00 & 0.35 \\
Qwen     & 0.39 & 0.32 & 0.47 & 0.38 & 0.35 & 1.00 \\
\bottomrule
\end{tabular}

\vspace{18pt}

\caption{Complete probe vocabulary. The geometric and irrelevant tiers are shared across all ten categories. Kansei axes are category-specific; the three shared-core axes (\textit{modern--traditional}, \textit{elegant--messy}, \textit{luxurious--cheap}) appear in every category, while the remaining axes are drawn per category from the Kansei literature and the LLM-assisted expansion described in Section~3.4.}
\Description{Table listing all geometric, irrelevant, and per-category affective adjective pairs used as probes.}
\label{tab:vocab}
\footnotesize
\setlength{\tabcolsep}{5pt}
\begin{tabularx}{\textwidth}{@{}lX@{}}
\toprule
\multicolumn{2}{@{}l}{\textbf{Geometric (all categories)}} \\
\multicolumn{2}{@{}X}{tall--short, boxy--curvy, simple--complex, elongated--compact, thin--thick, wide--narrow, angular--rounded} \\
\midrule
\multicolumn{2}{@{}l}{\textbf{Irrelevant (all categories)}} \\
\multicolumn{2}{@{}X}{loud--quiet, sweet--sour, hot--cold, musical--silent, edible--poisonous, polite--rude, wet--dry, scarce--abundant, juicy--stale, crowded--empty, public--private, fragrant--odorless, honest--deceptive, savory--bland, punctual--tardy, frequent--rare, famous--obscure, fluent--hesitant, active--passive, optimistic--pessimistic} \\
\midrule
\multicolumn{2}{@{}l}{\textbf{Affective (category-specific)}} \\
\midrule
Chair & modern--traditional, elegant--messy, luxurious--cheap, minimalist--ornate, sleek--bulky, plush--rigid, stable--wobbly, delicate--robust, graceful--clumsy \\
\addlinespace
Table & modern--traditional, elegant--messy, luxurious--cheap, opulent--austere, massive--airy, stable--wobbly, minimalist--ornate \\
\addlinespace
Lamp & modern--traditional, elegant--messy, luxurious--cheap, decorative--practical, graceful--clumsy, futuristic--vintage, minimalist--ornate \\
\addlinespace
Sofa & modern--traditional, elegant--messy, luxurious--cheap, plush--rigid, sturdy--fragile, spacious--compact, delicate--robust, sleek--bulky \\
\addlinespace
Cabinet & modern--traditional, elegant--messy, luxurious--cheap, minimalist--ornate, spacious--cramped, industrial--domestic, smooth--rugged, stable--wobbly \\
\addlinespace
Bookshelf & modern--traditional, elegant--messy, luxurious--cheap, minimalist--ornate, sturdy--fragile, spacious--cramped, delicate--robust, stable--wobbly \\
\addlinespace
Bottle & modern--traditional, elegant--messy, luxurious--cheap, minimalist--ornate, relaxed--sporty, lightweight--heavy, natural--artificial, eco-friendly--wasteful \\
\addlinespace
Jar & modern--traditional, elegant--messy, luxurious--cheap, minimalist--ornate, classic--trendy, futuristic--retro, disposable--lasting, delicate--robust \\
\addlinespace
Clock & modern--traditional, elegant--messy, luxurious--cheap, minimalist--ornate, robust--fragile, stylish--dowdy, creative--conventional \\
\addlinespace
Car & modern--traditional, elegant--messy, luxurious--cheap, sporty--sedate, masculine--feminine, smooth--rugged, aggressive--friendly, formal--casual \\
\bottomrule
\end{tabularx}
\end{table*}
\end{document}